\documentclass[letterpaper]{jpconf}
\usepackage{graphicx}
\begin{document}
\title{Adsorption properties and third sound propagation in superfluid $^4$He 
films on carbon nanotubes}

\author{Sonny Vo, Hossein Fard, Anshul Kogar, and Gary A. Williams}

\address{Department of Physics \& Astronomy, University of California,
Los Angeles, CA 90095}

\ead{gaw@ucla.edu}

\begin{abstract}
We consider the adsorption properties of superfluid 
$^4$He films on carbon nanotubes.  One major factor in the adsorption 
is the surface tension force arising from the very small diameter of the nanotubes.  
Calculations show that surface tension keeps the film thickness on the tubes very 
thin even when the helium vapor is increased to the saturated pressure.
The weakened Van der Waals force due to the cylindrical geometry also 
contributes to this.  Both of these effects act to lower the predicted velocity 
of third sound propagation along the tubes.  It does not appear that 
superfluidity will be possible on single-walled nanotubes of diameter about 
one nm, since the film thickness is less than 3 atomic layers even at saturation.  
Superfluidity is possible on larger-diameter nanotube bundles and 
multi-walled nanotubes, however.  We have observed third sound signals on
nanotube bundles of average diameter 5 nm which are sprayed onto a 
Plexiglass surface, forming a network of tubes.  
 \end{abstract}
 
\section{Introduction}
There has been considerable work exploring the nature of thin superfluid $^4$He films adsorbed on the surface of porous materials \cite{kotsubo,cho,rosenbaum,reppy,chan}.  Of particular interest is the superfluid phase transition, which in packed-powder porous materials has been shown to be a crossover from a 2D Kosterlitz-Thouless (KT) phase transition to a 3D vortex-loop transition.  The crossover occurs when the KT vortex pairs separate to distances larger than the pore size of the material, leading to a finite-size broadening of the transition \cite{cho}, and then at larger scales the formation of topological loop structures that lead to full 3D critical behavior \cite{williams}.  A different type of crossover has been predicted \cite{machta} for the case of a substrate consisting of long cylindrical tubes, a crossover from 2D to 1D.  Since a 1D system cannot support superfluidity at finite temperatures, the crossover will involve a pronounced loss of the superfluid fraction.  The parameter governing the crossover is the ratio of the vortex core size to the diameter of the cylinder;  the 1D limit will occur when this ratio is of order unity.  In a film whose superfluid thickness is less than a monolayer it has been found that the core size increases as the film is thinned \cite{cho}, so for a small enough cylinder size it may be possible to observe the crossover experimentally \cite{chu}.

The very small diameter of carbon nanotubes may allow a probe of this 1D limit.  Single-wall nanotubes have a diameter just over 1 nm, but it is fairly difficult to form a substrate of separated tubes, since they tend to agglomerate into bundles of larger diameter.  In addition, single tubes may not be able to support helium films thick enough for superfluidity, due to the very strong surface tension forces from the high curvature of the surface, a factor that we calculate in this paper.  For the first measurements of superfluidity on nanotubes we have employed a substrate of nanotube bundles of average diameter 5 nm.  We have successfully observed third sound propagation in this system, and report initial studies as a function of helium coverage at 1.3 K.

\section{Adsorption properties}

Previous third sound and torsion oscillator studies have been carried out on various graphite surfaces.  For a flat substrate the potential energy between a helium atom a distance $z$ from a  flat graphite substrate is given by $U(z) = - \alpha / z^3$, with $\alpha$ = 45 $K$ $layers^3$ where 1 layer is taken to be 3.6 \AA .  On a cylindrical substrate this expression is modified because some fraction of the substrate atoms are farther away from the helium atom.  The modified expression is given by \cite{cyl}
\begin{equation}
U(z) = -\frac{{18\pi {\kern 1pt} \alpha R^2 }}{{8\,(R + z)^5 }}F_1^2 \left[ {\frac{5}{2},\frac{5}{2};2;\frac{{R^2 }}{{(R + z)^2 }}} \right]
\end{equation}
where $F_1^2$ is Gauss' hypergeometric function.  Figure\,1 shows a comparison of this potential with that of the flat substrate for a 5 nm diameter cylinder.  For film thicknesses of order three layers there is about a 30\% reduction of the potential due to the geometric effect, and this will result in a thinner film than on the flat substrate.
\begin{figure}[h]
\begin{minipage}{14pc}
\includegraphics[width=18pc]{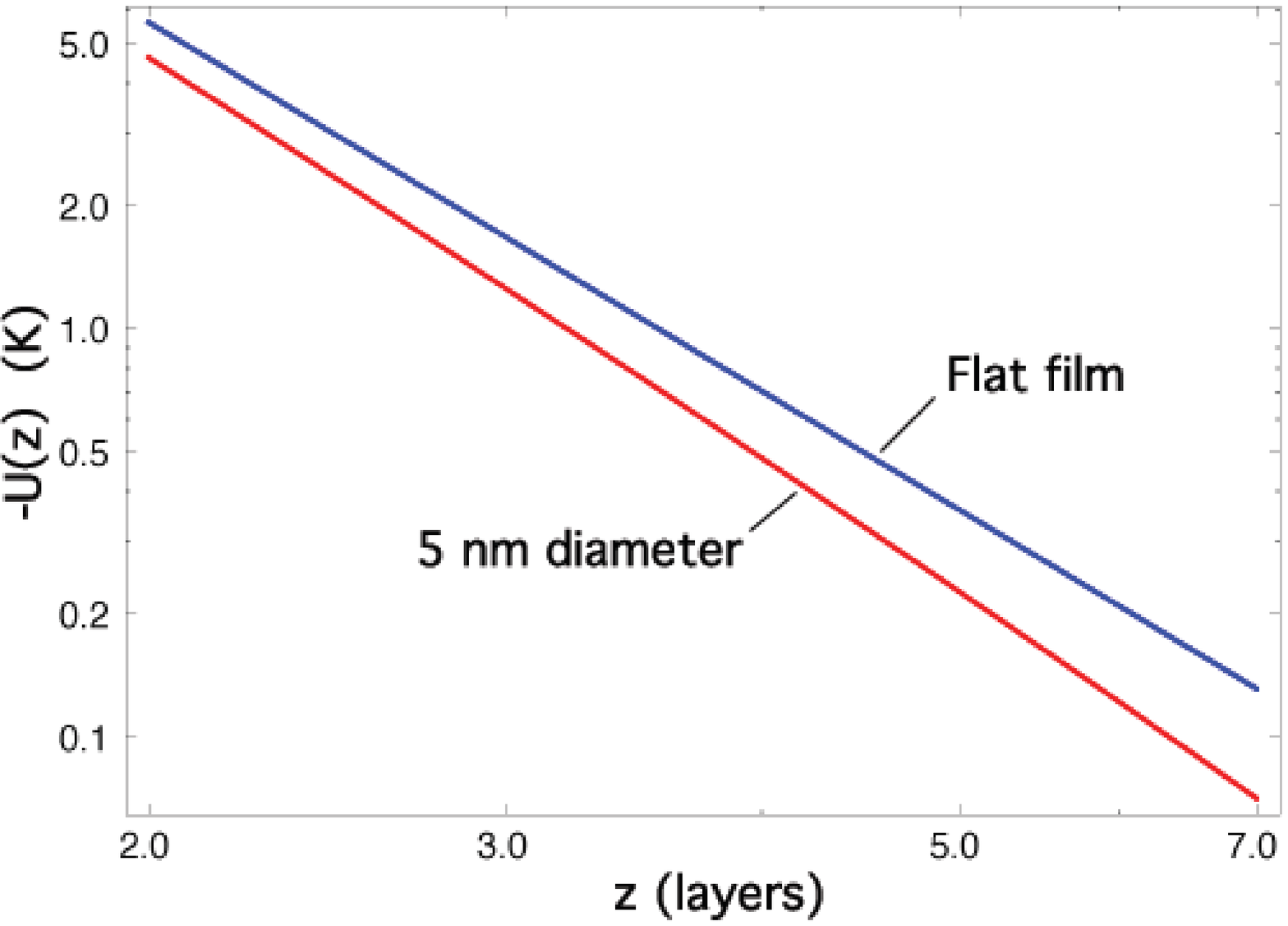}
\caption{\label{label}Van der Waals potential for a cylindrical substrate of 5 nm diameter compared to the potential for a flat substrate.}
\end{minipage}\hspace{5pc}
\begin{minipage}{14pc}
\includegraphics[width=20pc]{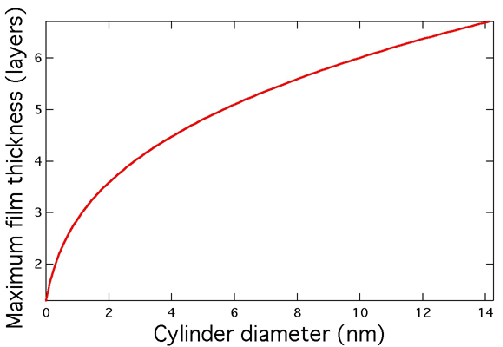}
\caption{\label{label}Film thickness at the saturated vapor pressure for a film on a cylindrical substrate as a function of the diameter.}
\end{minipage} 
\end{figure}

Another factor affecting the adsorption properties is the strong surface tension force resulting from the high curvature of the nanotubes.  Including the surface tension in the chemical potential of the film  gives a relationship between the film thickness $d$ and the ratio of the vapor pressure in the cell $p$ to the saturated vapor pressure $p_0$,
\begin{equation}
k_B T\ln \left( {\frac{{p_o }}{p}} \right) =  - U(d) - \frac{\sigma }{{\rho (R + d)}}
\end{equation}
This is similar to the result of Saam and Cole \cite{saam}, except here there is a sign change because they considered a film on the inside surface of a cylinder.  This equation can be solved numerically to find the film thickness for a given value of the cell pressure and the temperature.  Due to the surface tension term in Eq.\,2, the film stays quite thin even as the cell pressure approaches the saturated value.  Fig.\,2 shows the limiting thickness as a function of the nanotube diameter as the pressure approaches the saturated value. Since at temperatures around 1.3 K the critical film thickness needed for the onset of superfluidity is about three layers, it is doubtful that superfluidity can occur there on a single nanotube of diameter 1 nm.  The critical film thickness does decrease at lower temperatures, but it is unclear if superfluidity is possible or not.

The third sound speed for a film of thickness $d$ on the outside surface of a  cylinder is given by
\begin{equation}
c_3^2  = \frac{1}{{n^2 }}\frac{{\rho _s }}{\rho }\frac{{d - D_o }}{d}\left( {\left. {\frac{{\partial U(z)}}{{\partial z}}} \right|_{z = d}  - \frac{\sigma }{{\rho (R + d)^2 }}} \right)\left( {\frac{{R + d}}{2}} \right)\left( {1 - \frac{{R^2 }}{{(R + d)^2 }}} \right)
\end{equation}
where $n$ is the index of refraction accounting for the propagation along the random array of tubes, and $D_o$ is the effective thickness of the film that is not superfluid.  

\section{Measurements}

The substrate we use consists of nanotube bundles that are dissolved in distilled water using less than a minute of sonication.  Atomic force microscope studies of similar preparations \cite{gruner} show a range of bundle diameters, with an average value of 5 nm for the case of minimal sonication time, and an average tube length of order 1-2 $\mu$m.  These tubes are also 
functionalized with COOH end groups to aid in the dissolving process.  A spray gun pressurized with nitrogen is used to spray the liquid onto a two smooth plexiglass surfaces that form the resonator portion of the third sound experiment.  Spraying lightly for about two hours yields a black layer estimated to be 10-15 microns thick, where the tubes are well attached to the Plexiglass (it is fairly difficult to scrape them off the surface).  The two nanotube surfaces are mounted facing each other, separated by a Kapton gasket 250 microns thick.  This forms a rectangular third sound resonator 4 cm long and 1 cm wide.  The detector at one end is an epoxied 200 ohm Allen-Bradley resistor with a very strong temperature coefficient; at 1.3 K it increases to about 70 Kohms.  The casing on the nanotube side is sanded off to expose the carbon element, and after spraying the tubes they are scraped off the element since they would act to short it out.  The heater at the other end of the cell is formed from two epoxied wires connected by the nanotubes in between them, with a rectangular space carved around the wires to isolate those tubes from the rest.  There is also an epoxied wire to ground the rest of the nanotubes, but this was found in later experiments not to make any difference in the crosstalk between the heater and detector, which is minimal.  
\begin{figure}[h]
\begin{minipage}{14pc}
\includegraphics[width=18pc]{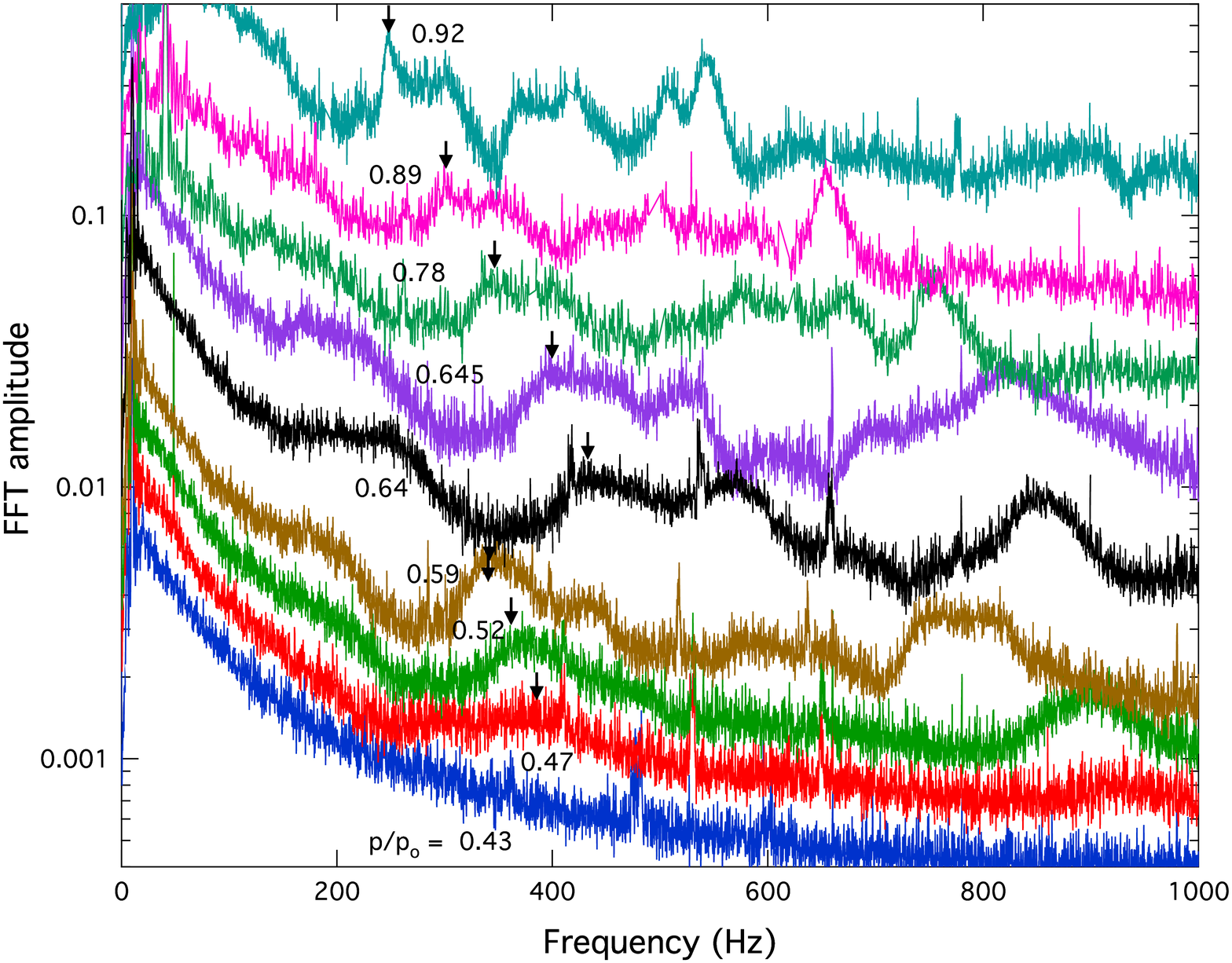}
\caption{\label{label}Averaged FFT amplitudes versus frequency, for a series of coverages with the indicated values of $p/p_o$.  The arrows indicate the second harmonic}
\end{minipage}\hspace{5pc}
\begin{minipage}{14pc}
\includegraphics[width=20pc]{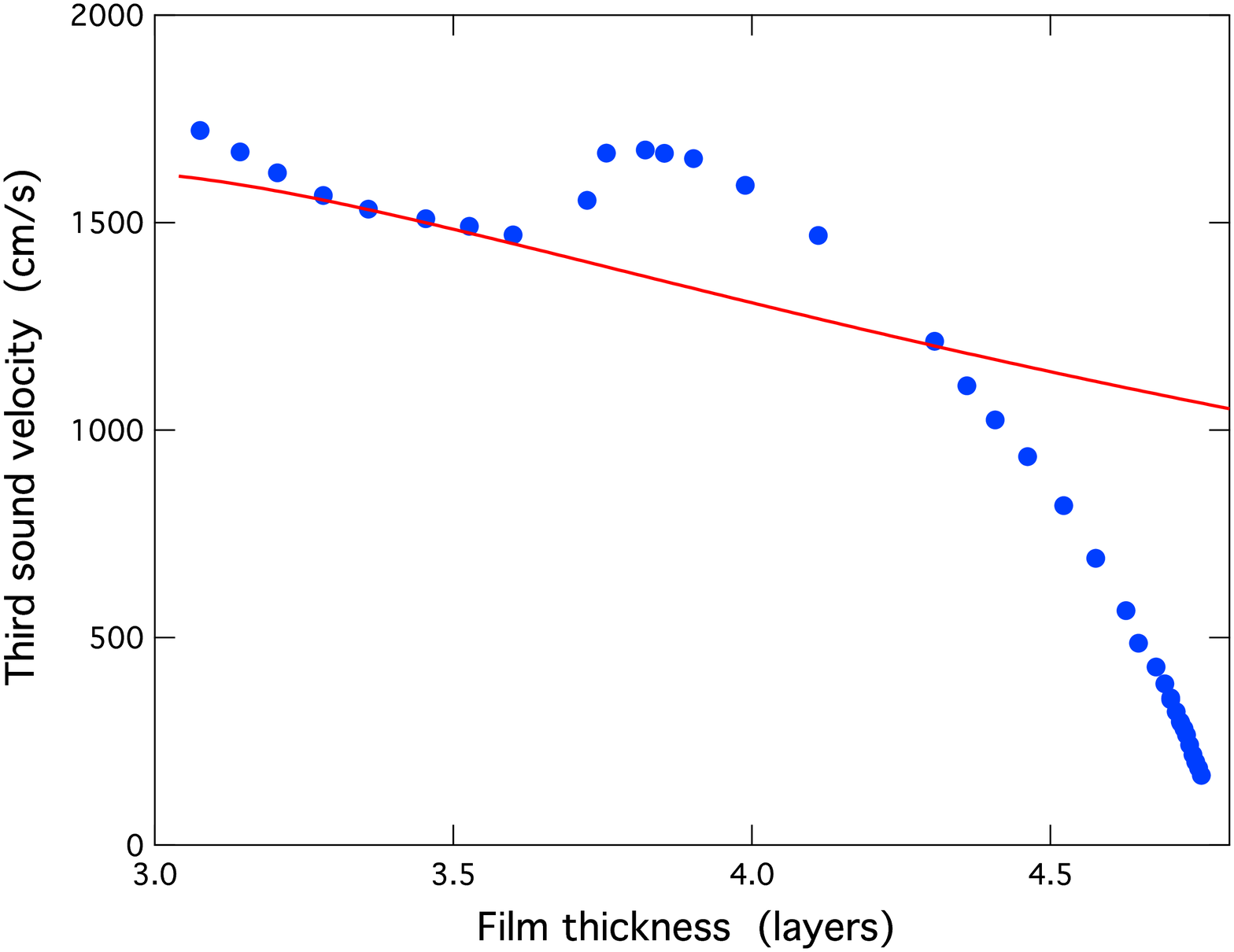}
\caption{\label{label}Third sound velocity at 1.3 K versus film thickness for $^4$He films adsorbed on 5 nm average diameter nanotube bundles.  The solid line is the theory, Eq.\,3}
\end{minipage} 
\end{figure}
The heater drive is swept in frequency from 5 Hz to 500 Hz with a 5 s repetition rate, with an rms power input of about 20 $\mu$W, and the response of the bolometer is monitored by biasing it with a 1 $\mu$a current from a battery and detecting the voltage oscillations with a preamplifier with a gain of about 5000.  The resulting signal is then analyzed with a LabView FFT module.  

Fig.\,3 shows a few of the averaged FFT curves for films at $T$ = 1.3 K with different values of $p/p_0$, where black arrows track the second harmonic.  The Kosterlitz-Thouless onset transition is apparent near $p/p_0$ = 0.47, where the modes are strongly broadened.  With added film the resonant frequencies decrease and sharpen up somewhat (though never better than Q = 10) until 
$p/p_0$ = 0.64, where there is a rapid increase in frequency, and then after that again a steady decrease.  Fig.\,4 shows the sound velocities deduced from these resonant modes, where the film thickness is computed from Eq.\,2.  The solid line is Eq.\,3, normalized to the data near 3.5 layers, and arbitrarily using D$_o$ = 2.2 layers.  We do not seem to observe any features connected with layering effects, as was seen in the measurements  on HOPG graphite surfaces near the superfluid onset \cite{chan}.  This may reflect inhomogeneities in the surface due to the bundling effect, which gives rise to a corrugated carbon surface.  

The increase in velocity near 
3.8 layers is history-dependent; raising the temperature of the cell from 1.3 to 1.5 K and back (partially evaporating and then recondensing the film) changed from the higher velocity back to the extrapolated curve of the thinner films, with a smaller increase then occurring above 4 layers.  It is likely that this increased velocity marks the onset of capillary condensation \cite{rosenbaum} at the points where the nanotube bundles touch.  The high curvature of the film at such points leads to an increased restoring force on the thickness waves.  In the presence of capillary condensation Eqs.\,2 and 3 are no longer valid, and this may account for the discrepancy between the data and the theoretical $c_3$ at the higher coverages.

\ack
This work is supported by the US National Science Foundation, Grant DMR 05-48521.  We thank Martti Kaempgen, David Hecht, and George Gruner for assistance in preparing the nanotubes, and Tim Hsieh, John Schulman, and Tyler Hill for help with the data collection.

\end{document}